\documentclass[superscriptaddress,preprint,amsmath,amssymb,aps,pra,longbibliography]{revtex4-1}

\setcitestyle{numbers,super}
\usepackage{fnpct}
\usepackage{placeins} 
\usepackage{upgreek}
\usepackage{amsmath} \usepackage{physics}
\usepackage{bm} \usepackage[margin=0.7in]{geometry}
\usepackage[colorinlistoftodos]{todonotes}
\usepackage[english]{babel}
\usepackage{fourier} 
\usepackage{array}
\usepackage{makecell}
\usepackage{tikz}
\usepackage{gensymb}
\usepackage{natbib}
\usepackage{dcolumn} 
\usepackage{graphicx}
\usepackage{epstopdf}

\begin{document}

\title{Wide-field Magnetic Field and Temperature Imaging using Nanoscale Quantum Sensors}
\author{Christopher Foy}
\affiliation{Department of Electrical Engineering and Computer Science, Massachusetts Institute of Technology, 50 Vassar St, Cambridge, MA, 02139, USA}
\affiliation{Research Laboratory of Electronics (RLE), Massachusetts Institute of Technology, 50 Vassar St, Cambridge, MA}
\author{Lenan Zhang}
\affiliation{Department of Mechanical Engineering, Massachusetts Institute of Technology, 77 Mass Ave, Cambridge, MA, 02139, USA}
\author{Matthew E. Trusheim}
\affiliation{Department of Electrical Engineering and Computer Science, Massachusetts Institute of Technology, 50 Vassar St, Cambridge, MA, 02139, USA}
\affiliation{Research Laboratory of Electronics (RLE), Massachusetts Institute of Technology, 50 Vassar St, Cambridge, MA}
\author{Kevin R. Bagnall}
\affiliation{Department of Mechanical Engineering, Massachusetts Institute of Technology, 77 Mass Ave, Cambridge, MA, 02139, USA}
\author{Michael Walsh}
\affiliation{Department of Electrical Engineering and Computer Science, Massachusetts Institute of Technology, 50 Vassar St, Cambridge, MA, 02139, USA}
\affiliation{Research Laboratory of Electronics (RLE), Massachusetts Institute of Technology, 50 Vassar St, Cambridge, MA}
\author{Evelyn N. Wang}
\email{enwang@mit.edu}
\affiliation{Department of Mechanical Engineering, Massachusetts Institute of Technology, 77 Mass Ave, Cambridge, MA, 02139, USA}
\author{Dirk R. Englund}
\email{englund@mit.edu}
\affiliation{Department of Electrical Engineering and Computer Science, Massachusetts Institute of Technology, 50 Vassar St, Cambridge, MA, 02139, USA}
\affiliation{Research Laboratory of Electronics (RLE), Massachusetts Institute of Technology, 50 Vassar St, Cambridge, MA}
\date{\today}

\begin{abstract}
The simultaneous imaging of magnetic fields and temperature (MT) is important in a range of applications, including studies of  carrier transport\cite{colliex_taking_2015,liao_photo-excited_2017,hu_spectral_2015}, solid-state material dynamics\cite{colliex_taking_2015,liao_photo-excited_2017,hassan_high-temporal-resolution_2017,hu_spectral_2015,nonnenmacher_scanning_1992,shi_scanning_2000}, and semiconductor device characterization\cite{pop_heat_2006,bagnall_simultaneous_2017}. Techniques exist for separately measuring temperature (e.g., infrared (IR) microscopy\cite{brites_thermometry_2012}, micro-Raman spectroscopy\cite{brites_thermometry_2012}, and thermo-reflectance microscopy\cite{brites_thermometry_2012}) and magnetic fields (e.g., scanning probe magnetic force microscopy\cite{hartmann_magnetic_1999} and superconducting quantum interference devices\cite{fagaly_superconducting_2006}). However, these techniques cannot measure magnetic fields and temperature simultaneously. Here, we use the exceptional temperature\cite{neumann_high-precision_2013, laraoui_imaging_2015} and magnetic field\cite{maze2008nanoscale,taylor2008high} sensitivity of nitrogen vacancy (NV) spins in conformally-coated nanodiamonds to realize simultaneous wide-field MT imaging.  Our "quantum conformally-attached thermo-magnetic" (Q-CAT) imaging enables (i) wide-field, high-frame-rate imaging (100 - 1000 Hz); (ii) high sensitivity; and (iii) compatibility with standard microscopes. We apply this technique to study the industrially important problem\cite{bagnall_simultaneous_2017,kuball_measurement_2002,sarua_integrated_2006,kuball_review_2016} of characterizing multifinger gallium nitride high-electron-mobility transistors (GaN HEMTs). We spatially and temporally resolve the electric current distribution and resulting temperature rise, elucidating functional device behavior at the microscopic level. The general applicability of Q-CAT imaging serves as an important tool for understanding complex MT phenomena in material science, device physics, and related fields. 
\end{abstract}

\maketitle

The NV center in diamond has attracted great interest because of its exceptional spin properties at room temperature, which exhibit outstanding nanoscale sensitivity to magnetic fields\cite{maze2008nanoscale,balasubramanian2008nanoscale,grinolds2014subnanometre,jensen2014cavity, glenn2015single,steinert_high_2010} and temperature\cite{neumann_high-precision_2013, laraoui_imaging_2015}. NV centers located within nanodiamonds (NVNDs) have gained particular interest for applications including drug delivery\cite{noauthor_biomedical_nodate}, thermal measurements of biological systems\cite{kucsko_nanometre-scale_2013,simpson_non-neurotoxic_2017,fu_characterization_2007,tanimoto_detection_2016, balasubramanian_nitrogen-vacancy_2014}, and scanning magnetometer tips\cite{tetienne_nature_2015,grinolds_nanoscale_2013}. The NVND's small size allows direct measurement of their local MT environment. These applications have motivated studies of NVND properties such as strain, magnetic and thermal sensitivity, and coherence time \cite{trusheim_scalable_2014,schirhagl_nitrogen-vacancy_2014}. However, NVND properties differ for a given fabrication process\cite{trusheim_scalable_2014} or surface treatment\cite{karaveli_modulation_2016}. This variability of nanodiamond material parameters and orientations has presented challenges for wide-field imaging studies using NVNDs. In this work, we (i) develop a model that describes the optically detected magnetic resonance (ODMR)\cite{noauthor_phys._nodate,gruber1997scanning} spectrum of NVND ensembles as a function of magnetic field and temperature; (ii) perform statistical characterization of NVND parameters, specifically the variation in NVND thermal response with implications for NVND temperature sensing; (iii) use this NVND model and our statistical characterization to extend the capabilities of NV sensing by enabling wide-field imaging with deposited coatings of NVNDs (Q-CAT imaging); (iv) demonstrate our technique's capabilities by imaging the dynamic phenomenon of electromigration; and (v) perform wide-field MT imaging of non-trivial commercial multifinger gallium nitride high-electron-mobility transistors (GaN HEMTs) and characterize the electric current distribution and the resulting temperature rise. 


\section{Sensing with NVND Ensembles}
\begin{figure*}[t!]
\centering
\includegraphics[scale=0.43]{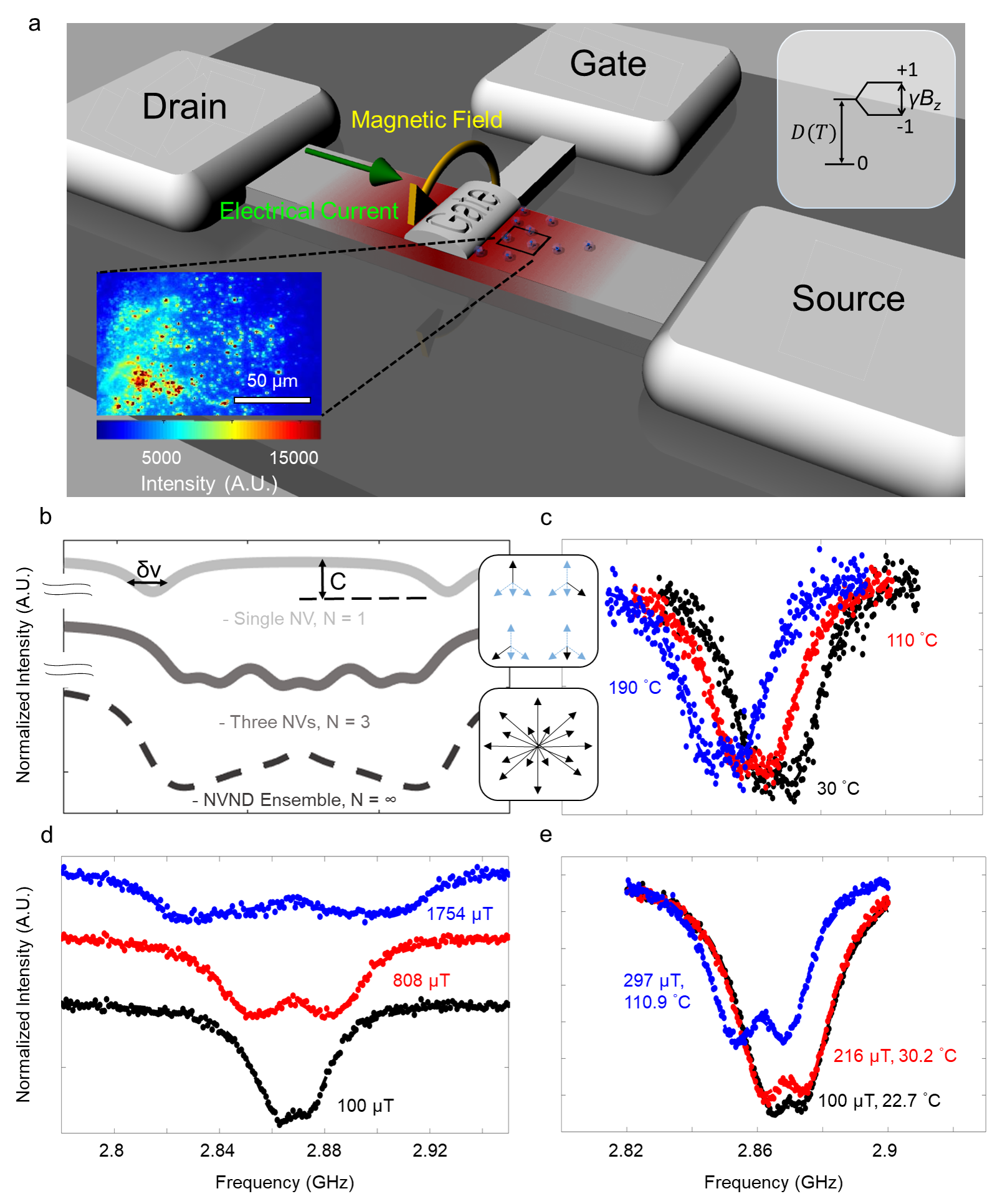}
\caption{\textbf{Overview of Q-CAT imaging}. \textbf{a}, Illustration of an application of Q-CAT imaging. A transistor experiences joule heating from electric current flowing between the drain and the source. (Inset bottom left), Fluorescence image of deposited NVNDs. NVNDs act as local probes of the magnetic field and temperature along the channel. (Inset top right), NV energy level diagram. \textbf{b}, Simulated OMDR spectra for the case of 1,3 and infinite NVs within the diffraction limit for a given \(\abs{B}\). Resonance seen in the ODMR spectra change are resolvable at low \(N\), but become inhomogeneously broadened as \(N \rightarrow \infty \). (inset), The top image shows NVs within defined crystallographic orientations such as those found in bulk diamond systems. The middle image shows 3 NV orientations constrained to the diamond crystallographic orientations. The bottom image represents NVs found in aggregated polycrystalline diamond. The NVs are modeled to be spherically symmetrical around an origin (bottom inset). \textbf{c}, Increase in temperature shifts \(D\) to lower frequencies. \textbf{d}, An increase in magnetic field broadens the ODMR curve. \textbf{e}, The different responses of the NV ODMR spectrum to \(\abs{\vec{B}}\) and T allows for simultaneous measurements of the NVND's MT environment.
}
\label{fig:figure1}
\end{figure*}

New diagnostic tools are needed to measure the MT profiles of microelectronics with high spatial and temporal resolution. For example, Fig:\ref{fig:figure1}a shows  the temperature and magnetic of a field-effect transistor, as illustrated in Fig:\ref{fig:figure1}a. As the gate electrode modulates source-drain current, magnetic fields (gold arrow) and temperature (red surface) vary with a geometry-dependent spatial distribution. High current densities at the electronic junction lead to high temperatures that accelerate device degradation. While numerous temperature mapping techniques are broadly used for characterization: infrared (IR) microscopy\cite{brites_thermometry_2012}, micro-Raman spectroscopy \cite{brites_thermometry_2012}, and thermo-reflectance microscopy\cite{brites_thermometry_2012}, each of these techniques has many undesirable features. IR and thermo-reflectance microscopy are both commonly used wide-field techniques but suffer from complicated calibration procedures and difficulties in measuring adjacent metal and semiconductor regions at the sample surface. Micro-Raman spectroscopy has the major advantage of directly probing the temperature in the active device layers, but is limited to serial acquisition with a \(\sim\)1 \(\upmu\)m diameter spot. Further, even if the temperature profile could be captured by the above techniques, the electric/magnetic current distribution would still not be understood, or should be measured separately. A wide-field MT measurement technique with simple instrumentation and calibration procedures would therefore be of great value for electronics characterizations. To address these requirements we introduce a new method that takes advantage of a film of NVNDs for nanoscale spin-based magnetometry and thermometry enabling Q-CAT imaging (Fig:\ref{fig:figure1}a inset). To generate images using Q-CAT imaging, we first determine the local magnetic field and temperature across a single NV by monitoring its average fluorescence intensity.  In the \(\ket{m_s = 0}\) "bright" state, the NV is photostable; in contrast, in the \(\ket{m_s = \pm1}\) state, the NV undergoes an intersystem crossing into a metastable state\cite{noauthor_phys._nodate,doherty_nitrogen-vacancy_2013}. This decay path is non-radiative in the visible spectrum and thus reduces the NV's average fluorescence intensity. Fluctuations in the NV's local thermal environment change the \(\ket{m_s = 0}\) \(\rightarrow\) \(\ket{m_s = \pm1}\) zero field splitting parameter, \textit{D(T)}\cite{acosta2010temperature}, whereas magnetic fields lift the degeneracy of the NV's \(\ket{m_s = \pm1}\)  state through the Zeeman effect\cite{taylor2008high}. For weak non-axial or aligned magnetic fields (\(\vec{B}\) \(\sim\) \(B_z\) \(<\) 100 mT), the resonance frequencies of the \(\ket{m_s = \pm1}\) states are given by\cite{rondin_magnetometry_2013}
\begin{equation}\label{equation1}
\epsilon(T,B_z)_{\pm} = D(T) \pm{\sqrt{E^2 +(\gamma B_z)^2}}
\end{equation}
where \(\epsilon_{\pm}\) are the NV resonance frequencies corresponding to the \(\ket{m_s = 0}\) \(\rightarrow\) \(\ket{m_s = \pm1}\) transisitions, \(\gamma\) is the NV gyromagnetic ratio, \(E\) is the NV's off-axis zero-field splitting parameter\cite{rondin_magnetometry_2013} which results from local strain, and \(B_z\) is the axial magnetic field. This behavior of the \(\ket{m_s = \pm1}\) states with magnetic field and temperature is represented graphically in Fig:\ref{fig:figure1}a inset top.
We can measure \(\epsilon_{\pm}\) through ODMR experiments(Fig:\ref{fig:figure1}b-e)\cite{noauthor_phys._nodate} which allows us to determine \textit{D(T)} and \(B_z\). 

Fig:\ref{fig:figure1}b (top line) shows an ODMR spectrum of a single NV with an applied \(B_z\). This spectrum has the form
\begin{equation}\label{equation2}
I(f;\epsilon ,\delta\nu,C) = 1 - 1/2 \sum_{m = 1}^{2}{L(f;\epsilon_m,\delta\nu,C)}
\end{equation}
where \(L(f;\epsilon,\delta\nu,C)\) is the three-parameter Lorentzian function, \(f\) is the microwave frequency, \(\delta\nu\) is the linewidth of the transition, and  \(C\) is the change in fluorescence rate between the \(\ket{m_s = 0}\) and \(\ket{m_s = \pm1}\) states.   

As the number of distinct NV center orientations, \(N\), for a given \(B_z\) within a diffraction-limited volume increases, the final observed ODMR spectrum is the sum of the individual NV ODMR spectra having the form 
\begin{equation}\label{equation3}
I(f;\epsilon ,\delta\nu,C_m) = 1 - 1/(2N)\sum_{i = 1}^{N}{\sum_{m = 1}^{2}{L(f;\epsilon_{mi},\delta\nu,C)}}
\end{equation}
as seen in Fig: \ref{fig:figure1}b (middle line) for $N = 3$. NV centers located in single crystal bulk diamond can have at most 4 NV orientations\cite{maertz2010vector,wang2015highvector, clevenson2018robust, schloss2018simultaneous}, belonging to the set shown in Fig:\ref{fig:figure1} inset top, which lie along the diamond's [111] crystallographic axes.

For NVND ensembles, the number of possible NV orientations is not fixed\cite{chen_wide-field_2013}. As the number increases, NV resonances are no longer individually resolvable, as shown in Fig: \ref{fig:figure1}b (bottom line). To describe this behavior, we model the NVND ensemble as an isotropic distribution of NV orientations, each oriented at an angle \(\theta \in [0,\pi]\) with respect to the objective axis (inset Fig: \ref{fig:figure1}b bottom). The ODMR spectrum becomes
\begin{equation}\label{equation4}
1 -I(f;T,\abs{B},\delta\nu,C) \propto \int_{0}^{\pi}{\sum_{m = 1}^{2}{R(\theta)L(f;\epsilon (T,\abs{B})_{m\theta},\delta\nu,C)}\sin{\theta} d\theta}
\end{equation}
where the summation becomes an integral, and \(R(\theta)\) represents the angular dependent weighting of the NV's normalized  fluorescence value (see Supplemental Information (SI)). The NVND ODMR spectrum as a function of magnetic field and temperature is shown in Figs: \ref{fig:figure1}c-e. Fig: \ref{fig:figure1}c illustrates how the center of the ODMR spectrum shifts to lower frequencies as its ambient temperature increases. Due to the isotropic model, NVNDs are insensitive to magnetic field orientation, so \(B_z\) \(\rightarrow\) \(\abs{\vec{B}}\). Fig: \ref{fig:figure1}d demonstrates how NVND ODMR spectra broaden as a function of \(\abs{B}\). Finally, Fig:\ref{fig:figure1}e shows \(I(f;D(T),\abs{B},\delta\nu,C)\) for  two values of \(\abs{B}\) and \(T\) (see SI for complete derivation). Fitting the measured spectra to Equation \ref{equation4} allows the determination of both \(\abs{B}\) and \(D(T)\). 

\section{Statistical Characterization of NVND Ensembles}

Temperature imaging with NVNDs requires extracting \(T\) from \(D(T)\). Thus, \(D\) as a function of temperature needs to be determined. Previous studies have investigated this dependence with limited numbers of NVNDs\cite{plakhotnik2014all}, degrading the precision and accuracy of \(D(T)\). In addition, NVNDs can have different \(D(T)\), originating from varying nanodiamond impurity concentration, strain, and surface geometry\cite{acosta2010temperature,doherty_temperature_2014}. To determine \(D(T)\) we calibrate Q-CAT imaging with known temperatures. Importantly, the wide-field nature of Q-CAT imaging allows for the investigation of several hundreds of nanodiamonds in parallel - ideal for studying nanodiamond properties. This allows for determination of the NVND's \(D(T)\) distribution.

\begin{figure*}[b!]
\centering
\includegraphics[scale=0.55]{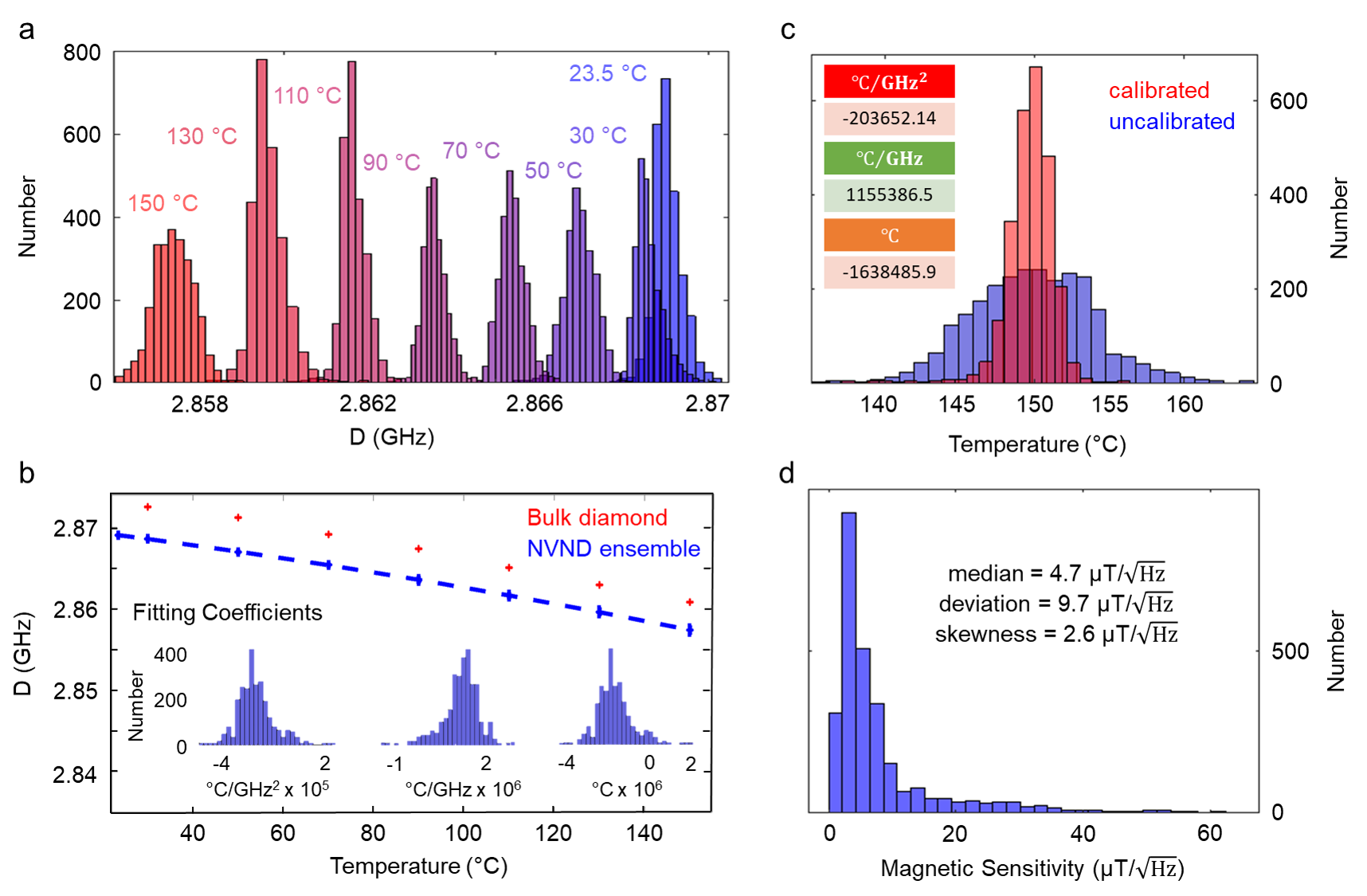}
\caption{\textbf{Statistical analysis of NVND properties}. \textbf{a}, Histograms of \(D\) as a function of T for NVND ensembles. \textbf{b}, Fit of the mean value of \(D\) vs T. (inset), Histogram of the fit coefficient. \textbf{c}, Histogram of the measured temperature across all NVNDs using the average fit coefficients (uncalibrated) and individual fit coefficients for each NVND (calibrated). \textbf{d}, Distribution of NVND magnetic sensitivity.
}
\label{fig:figure2}
\end{figure*}

To determine \(D\) as a function of temperature, we tracked \(D(T)\) across 2573 commercially available 100-nm CO-OH terminated nanodiamonds (Ad\'amas Nanotechnologies) from 23.5-150 $^{\circ}$C (Fig: \ref{fig:figure2}a) using a temperature controlled stage (See methods). Fig:\ref{fig:figure2}b shows a second-order polynomial fit of the average NVND \(D(T)\) compared with the measured \(D(T)\) of bulk diamond NVs. NVNDs' \(D\) is lower than bulk diamonds across all temperatures, which we attribute to the nanodiamond's large strain (E $>$ 5 MHz). Histograms of the distribution of fitting coefficients are within Fig: \ref{fig:figure2}b inset. We note that the mean measured standard deviation of the NVND thermal response across all temperatures (3.88 $^{\circ}$C) is greater than what is expected from measurement error (2.17 $^{\circ}$C \(\pm\) 0.46 $^{\circ}$C), indicating that individual NVNDs have varying thermal responses. This variation could result from differences in the thermal resistance at the nanodiamond sample interface, or, as previously theorized\cite{acosta2010temperature}, inherent variation in NVND thermal response. Fig:\ref{fig:figure2}c shows the distribution in the measured temperature using the mean value of the fitting coefficients (Fig:\ref{fig:figure2}c inset). While individual NVNDs have shot-noise scaling in determining \(D\) at a particular temperature, Fig:\ref{fig:figure2}c demonstrates that the variation in thermal response among the NVND ensemble limits their temperature accuracy for imaging applications.


To compensate for the variation in \(D(T)\), we apply the fitting coefficients in Fig: \ref{fig:figure2}b inset to each NVND. This individual NVND \(D(T)\) calibration improves the mean temperature precision to 2.6 $^{\circ}$C per pixel ($\sim$ 1 \(\upmu\)m x 1 \(\upmu\)m) across all NVNDs, which is within the variation expected from the experimental noise. These results suggest that improving the uniformity of \(D(T)\) among NVNDs should be a priority for future nanodiamond fabrication studies. The precision of Q-CAT imaging is comparable to that of alternative thermal imaging techniques such as micro-Raman spectroscopy (1 $^{\circ}$C)\cite{brites_thermometry_2012} and IR (1 $^{\circ}$C) \cite{brites_thermometry_2012}. 

We also investigated the NVND's magnetic field sensitivity (Fig:\ref{fig:figure2}d) and measured a median of 4.7 \(\upmu\)T\(/\sqrt{\textrm{Hz}}\) (see SI). We measure a standard deviation of 9.7 \(\upmu\)T\(/\sqrt{\textrm{Hz}}\) and a skewness of 2.6 \(\upmu\)T\(/\sqrt{\textrm{Hz}}\). This distribution contains values ranging from 1.2 to 64.4 \(\upmu\)T\(/\sqrt{\textrm{Hz}}\). This range indicates that future studies on NVND sensitivity need to carefully consider NVND median sensitivity and the entire distribution in order to rigorously determine how changes to NVND surface chemistry\cite{nagl_improving_2015} and fabrication affect performance. 

\section{Demonstrating Q-CAT Imaging}

\begin{figure*}[!ht]
\centering
\includegraphics[scale=0.55]{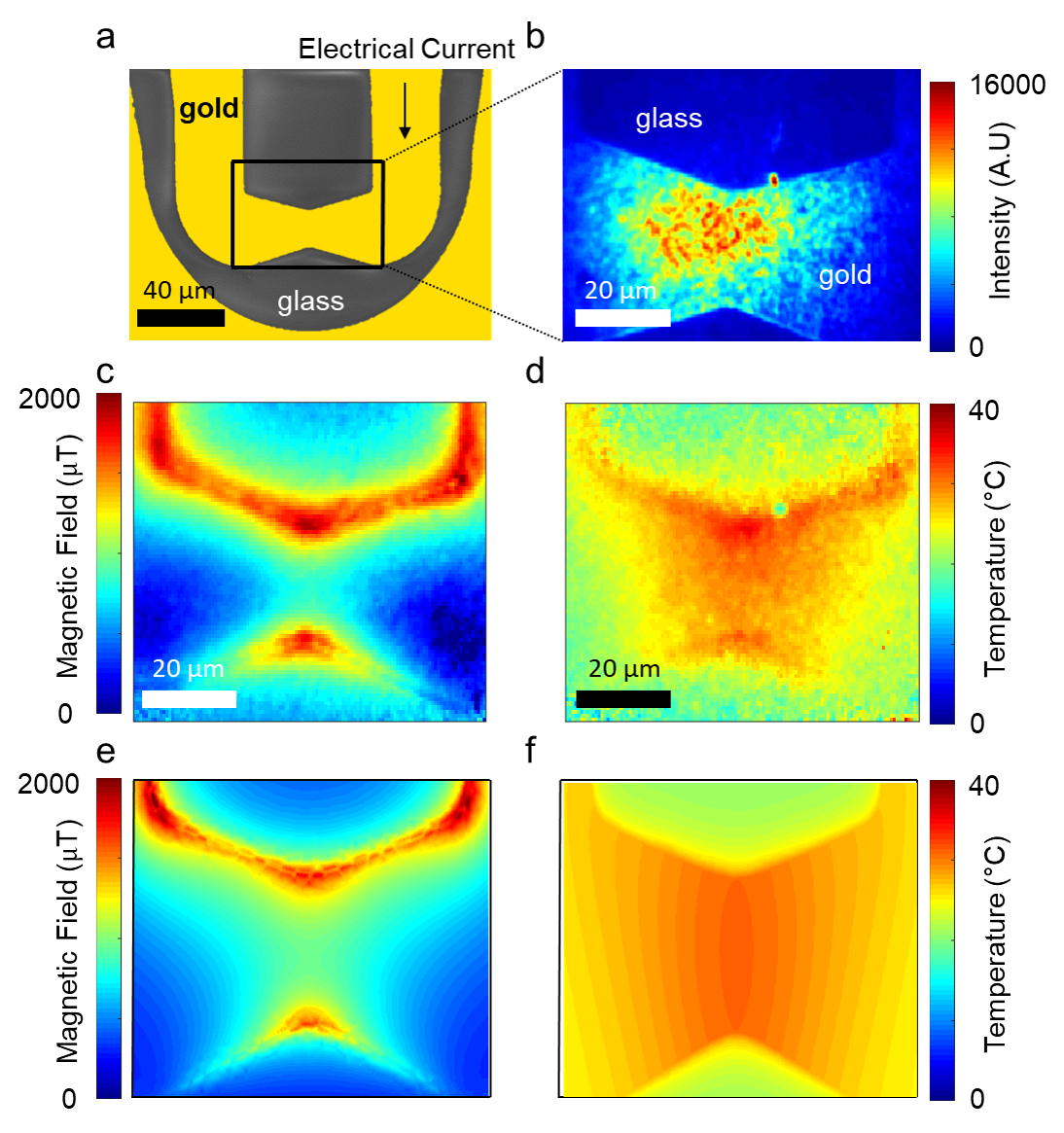}
\caption{\textbf{Q-CAT imaging of microfabricated structure}. \textbf{a}, False-color SEM of microfabricated structure to identify device structure. \textbf{b}, Fluorescence image of deposited NVNDs. \textbf{c-d}, Magnetic and temperature image of the ROI indicated in a. \textbf{e-f}, MT simulations of the tapered region.
}
\label{fig:figure3}
\end{figure*}

To demonstrate Q-CAT imaging, we image one of the simplest structures that couples electromagnetism and heat transport: a thin gold wire with a narrowed tapered region. Fig: \ref{fig:figure3}a is a scanning electron microscope (SEM) image of the test system. Electric current (30 mA) induces significant local joule heating around the tapered region (dashed black box). The surrounding metal structure supplies the resonant microwave field, which drives the NV's spin resonances. We deposit NVNDs on top of the structure (thickness $\sim$ 100 nm, see SI). This NVND coating solves two issues found with bulk diamond NV thermometry; due to the conformal coating it has good thermal contact with substrates (see SI), and avoids heat spreading due to the much lower thermal conductivity of diamond nanoparticles\cite{noauthor_relative_nodate}. Both issues limit spatial resolution and artificially reduce peak temperature (see SI). An electron multiplying charge coupled device (EMCCD) camera images the NVND red fluorescence as shown in Fig: \ref{fig:figure3}b. Fig:\ref{fig:figure3}c-d are images of the magnetic field and temperature of the tapered region, which are measured simultaneously by extracting \(\abs{B}\) and \(D\) from fits of the ODMR spectrum measured at each pixel. We determine temperature by converting \(D\) using the coefficients presented in Fig: \ref{fig:figure2}c inset. These images give insight into how the current flows through the structure, with a high magnetic field region at the structure's edge because the current density increases from both the structure's tapering and magnetic contributions from the side walls. We also see that the temperature is 10 $^{\circ}$C higher at the kinked region, which indicates that this area is the probable point of failure for the device. Fig:\ref{fig:figure3}e-f show MT (COMSOL) simulations of the taper region. We note strong agreement between measurements and simulations, and we attribute deviations from the simulations to the unknown fluctuations in the structure's surface morphology. According to this proof-of-concept study, two significant advantages of the proposed technique have been clearly shown. First, the electric/magnetic current distribution and the associated heating effect has been demonstrated at the wide-field. Second, the microscale temperature distribution on the glass substrate, which has a weak thermo-reflectance and Raman signature, is well-resolved through Q-CAT imaging.

\section{Q-CAT Imaging of Dynamic Processes}

While the previous experiment was at steady state, we also we wish to showcase Q-CAT imaging of a dynamic process which is difficult to capture with conventional techniques\cite{brites_thermometry_2012}. Thus, we apply Q-CAT imaging to study electromigration within microstructures.  Electromigration is a runaway process, which concerns the failure of a conductor due to momentum transfer between the conducting electrons and the metal atoms, causing the metal ions to move and creating discontinuities. As we will show, a wide-field technique like Q-CAT imaging is well-suited to study this dynamic phenomenon because it allows high frame-rate-MT imaging. 

\begin{figure*}[t!]
\centering
\includegraphics[scale=0.4]{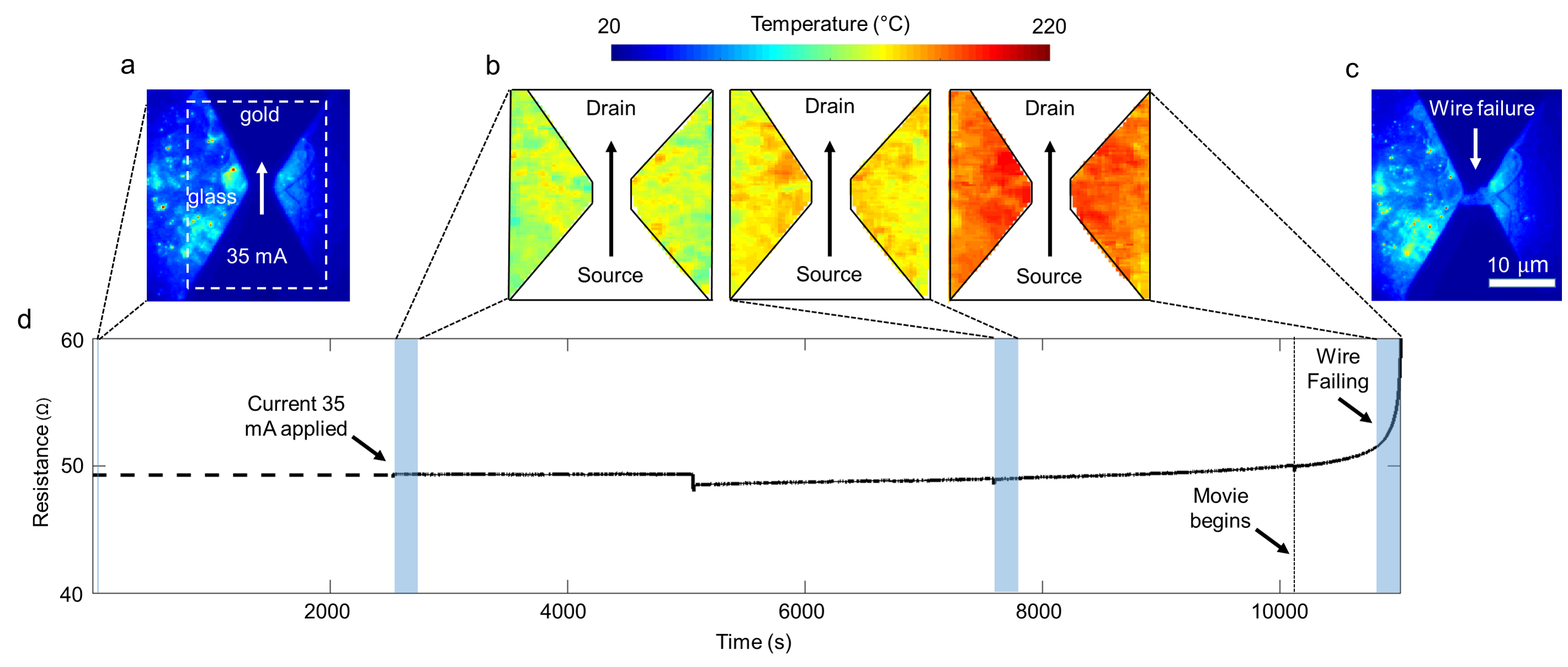}
\caption{\textbf{Q-CAT imaging of electromigration}. \textbf{a}, Fluorescence image of kinked wire after NVND deposition. \textbf{b}, Temperature images of the wire, after 35 mA of current is applied, until wire failure. \textbf{c}, Fluorescence image of wire after failure. \textbf{d}, Resistance as a function of time. Resistance was determined by dividing the applied voltage by the current. Boxes represent measurement time for each image. The periodic jumps are artifacts that result from the experiment being suspended as the camera buffer is emptied. The dashed line is the extrapolated resistance before current was applied. The attached video's frame rate was artificially increased to shorten video time. 
}
\label{fig:figure4}
\end{figure*}

To begin the electromigration process, we increase the current through a similar kinked wire to a constant 35 mA and drive the device to failure. Fig: \ref{fig:figure4}a is a fluorescence image of the structure at t = 0 s. Fluorescent areas are deposited NVNDs.  Q-CAT imaging shows that the temperature increases over three distinct time intervals as the wire undergoes electromigration (Fig: \ref{fig:figure4}b). In the first shaded region, the device is still operating normally, and resistance is constant. In the second shaded interval, the resistance starts to linearly increase and in the final region, failure is imminent and the resistance of the wire exponentially increases (Fig: \ref{fig:figure4}d). The local temperature increases over time from 50 $^{\circ}$C to 220 $^{\circ}$C. Within 10 ms of the final image shown in Fig: \ref{fig:figure4}b, the wire breaks and can no longer support electric current. Fig:\ref{fig:figure4}c is a fluorescence image of the wire after failure with a tear at the kink, the location of temperature maximum, which is the point of failure of the device. The high temperature indicates the location of maximum electric current density and highest electron-gold momentum transfer -- the source of electromigration. As expected, due to the constant current of 35 mA, the overall magnetic field distribution did not change over the course of the measurement. This process, beginning at the vertically dashed line, is shown in the attached video, which is sped up for ease of viewing. 


The exposure time for each fluorescence image was 10 ms, with a total integration time of 100 ms per microwave frequency, corresponding to a total measurement time for each of the shaded intervals of $\sim$48 seconds. However, 83\(\%\) of the measurement time was occupied by camera readout time. In addition, the high count rate of the NVNDs (< 1x10\(^7\) cps) coupled with the small electron well depth of our camera limited the minimum exposure time. For a camera optimized for bright samples, the exposure time could be decreased further, by increasing the optical pump power. This limitation indicates that a frame rate of 100-1000 Hz is possible if a sparse sampling scheme was adopted (see SI) and the experimental overhead frame readout time (50 ms) was eliminated. This frame-rate compares favorably to \textit{in situ} methods such as micro-Raman which requires $\sim$1 second per pixel, and is comparable to other wide-field techniques such as IR imaging or thermo-reflectance microscopy.


\section{Q-CAT imaging of G\MakeLowercase{a}n HEMTs}

We further expand Q-CAT imaging beyond proof of principal experiments. We will image a technologically important problem where the interplay between temperature and electric currents is crucial and Q-CAT's wide-field MT imaging capability is essential - GaN HEMTs.  GaN HEMTs are field-effect transistors which incorporate a junction between two materials with different band gaps. They are increasingly used in applications ranging from radio frequency amplifiers\cite{zanoni_algan/gan-based_2013} to high power electronics\cite{chowdhury_lateral_2013}. The extremely high power density ($>$ 5 W/mm) in GaN HEMTs gives rise to a concentrated ($\sim$1 $\upmu$m) channel temperature ($>$200 $^{\circ}$C), which leads to device failure\cite{bagnall_simultaneous_2017,kuball_measurement_2002,sarua_integrated_2006,kuball_review_2016}. The operation of GaN HEMTs involves non-linear coupling between electromagnetism and heat transport, making their dynamics difficult to capture by simulation, especially when considering device variations. The problem is particularly challenging for the complex geometries of commercial, multi-finger GaN HEMTs. We will image GaN's MT environment to both understand how current flows through the device and identify temperature maxima - the likely failure point, whose magnitude is predictive of device lifetime.

\begin{figure*}[t!]
\centering
\includegraphics[scale=0.55]{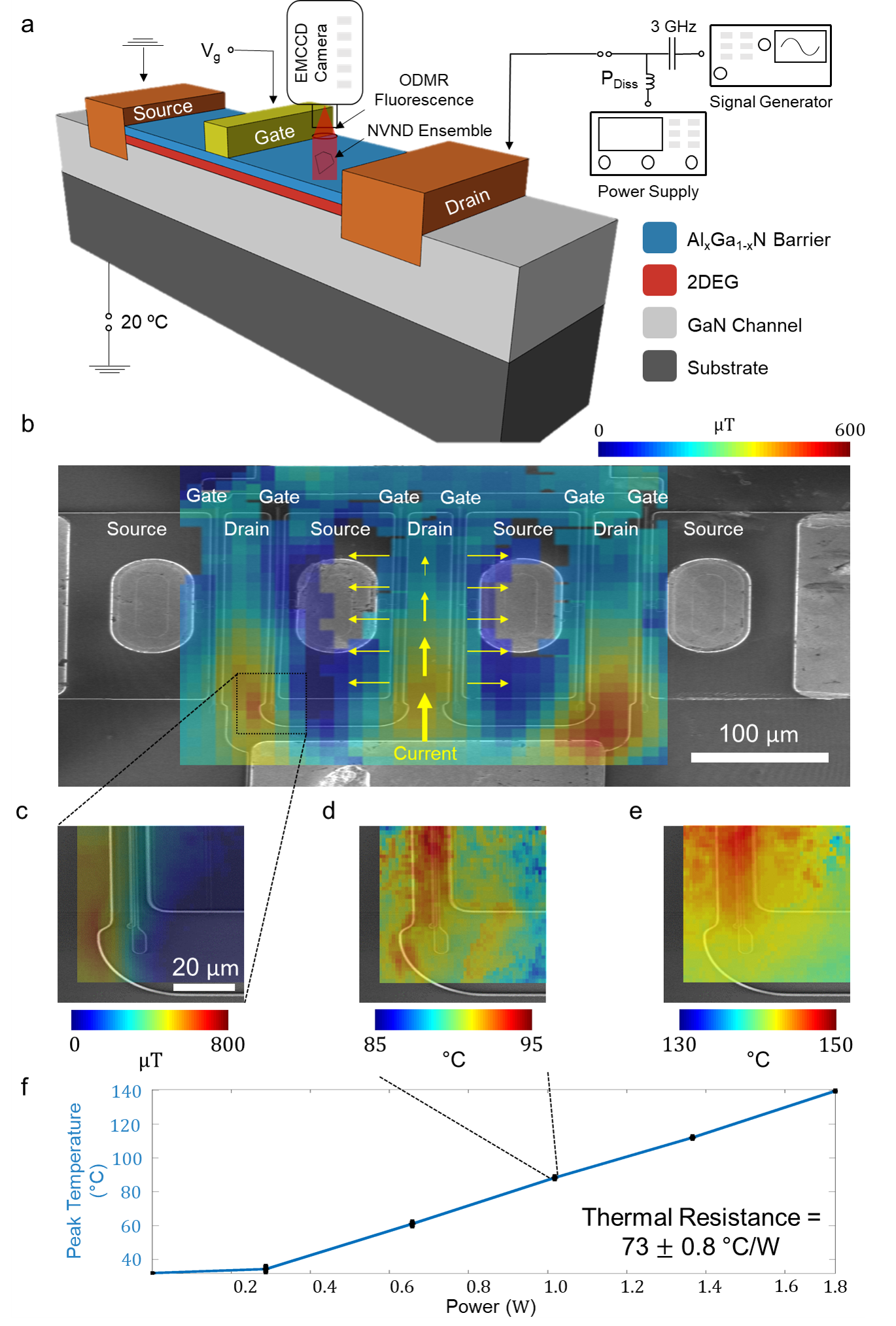}
\caption{\textbf{Q-CAT imaging of a multifinger GaN HEMT}. \textbf{a}, Schematic of Q-CAT imaging of a GaN HEMT. \textbf{b}, Magnetic image of a six-finger GaN HEMT in the ON state (4 V, 72 mA). An SEM is superimposed to guide the eye. The magnetic field concentrates at the drain and decreases along the channel width. \textbf{c}, High-resolution magnetic field (290 mW) and \textbf{d-e}, Temperature images of the channel stop (1 W and 1.73 W respectively). \textbf{f}, Peak temperature in the ON state as a function of the drain bias. We measure a thermal resistance of 73 \(\pm\) 0.8 $^{\circ}$C/W.}
\label{fig:figure5}
\end{figure*}

Fig:\ref{fig:figure5}a illustrates a GaN HEMT under Q-CAT imaging. The current, carried as a 2D electron gas (2DEG) from drain to source, is modulated by the top gate. Fig.\ref{fig:figure5}b shows the associated magnetic field for a commercial 6-finger GaN HEMT superimposed over an SEM of the device. A power of 290 mW was applied across the drain while a gate voltage of -2.5 V (threshold voltage -2.8 V) was applied to keep the device in the ON state.  The magnetic field decreases from the base of the drain by $\sim$ 300 \(\upmu\)T as the current drops along the channel width, as illustrated by the yellow arrows. 

We sought to investigate the area with high magnetic field, the channel stop, and thus we conducted high resolution (50x) MT imaging of that region (Fig.\ref{fig:figure5}c-e). As expected, the magnetic field of the left side of the channel stop is higher, because the resistance of the GaN channel is much higher than that of the drain. The resulting temperature profile is obtained simultaneously. Significant temperature rise is localized around the gate, which agrees with previous experimental observations \cite{bagnall_simultaneous_2017,kuball_measurement_2002,sarua_integrated_2006,kuball_review_2016}. Particularly, attributed to the high spatial resolution and wide-field nature of Q-CAT imaging, a sharp temperature drop is well-resolved at the end of the gate along the channel direction (Fig.\ref{fig:figure5}d-e), indicating limited leakage current at the channel's end. Fig.\ref{fig:figure5}f is the peak temperature as a function of dissipated power. We measure a thermal resistance of 73 \(\pm \) 0.8 $^{\circ}$C/W, which agrees with previous measurements of the same model(75 $^{\circ}$C/W)\cite{bagnall_simultaneous_2017}. These results elucidate device physics at a spatial resolution that is competitive with \textit{in situ} methods, at wide-field.

\section*{Discussion}

Significant improvements in the sensitivity of Q-CAT imaging are possible. They could be achieved by NVNDs possessing coherence times approaching what has been demonstrated with bulk diamond\cite{balasubramanian_ultralong_2009,trusheim_scalable_2014}. Sensitivity could be further extended by using dynamical decoupling pulse sequences\cite{taylor2008high}. We estimate that \(\textrm{mK}/ \sqrt{\textrm{Hz}}\) \cite{kucsko_nanometre-scale_2013} temperature precision and sub \(\textrm{nT}/\sqrt{\textrm{Hz}}\)
 \cite{balasubramanian_ultralong_2009}magnetic field sensitivity are achievable with Q-CAT imaging. Q-CAT imaging has further advantages not demonstrated in this work. The unique method of NVND deposition enables the imaging of non-planar geometries (see SI). Also, the temporal resolution of this technique could be extended for periodic signals through stroboscopic imaging, in which the laser readout for the NVNDs is pulsed in sync with the application of the electric current. In this manner, a temporal resolution of ~10 ns could be achieved, which is limited by the laser gating time. This application could be of interest in studying the transient behavior of the MT environment of microelectronic devices, such as the peak temperature of GaN HEMTs at MHz frequencies. 
 
In conclusion, Q-CAT imaging has a number of significant advantages: wide-field measurement with a field of view greater than 100 \(\upmu\)m x 100 \(\upmu\)m, compatibility with microscopes and almost all materials, and a mean thermal sensitivity comparable to micro-Raman spectroscopy and IR microscopy. Further, it has a diffraction limited spatial resolution ($<$ 1 $\upmu$m). One possible shortcoming of Q-CAT imaging is that it requires samples that are resilient to microwave fields, which is common for most solid-state devices.  In addition, because Q-CAT imaging is an optical measurement, it requires samples with low background red fluorescence (< 1x10\(^6\) cps) under green  excitation. We believe that these requirements are not especially restrictive for many active fields of research. Allowing, at a fundamental level, investigation of both steady state and transient thermal transport in a variety of materials systems and characterization of material thermophysical properties. At the device level, Q-CAT imaging helps understand the working principle, life-time and failure mechanisms of devices. These applications are of interest to various fields such as microelectronics, ceramic memories and lithium-ion batteries.

\section{Methods}
\subsection*{Experimental Setup}
A Verdi g2 532 nm single mode longitudinal laser was focused through a custom microscope onto the back aperture of an objective. The resulting collimated excitation beam was used to pump the NVNDs. Emitted red fluorescence from the NVNDs was collected and measured using an Andor EMCCD camera.  A 532 nm notch filter and 650 nm long pass were used to eliminate background fluorescence from the green excitation pump and the neutral charge state of the NV, respectively. The microwave excitation was swept by using a signal generator (Hewlett Packard ESG - D4000A). Collected fluorescence was correlated with microwave frequency in post processing to determine the ODMR spectra shown in the main text. 

\subsection*{Microfabricated Structure Preparation}
The sample shown in Fig:\ref{fig:figure3} and Fig:\ref{fig:figure4} were fabricated in MIT's cleanroom facility, the Microsystems Technology Laboratory, using photolithography. We deposited a positive resist (S1813) onto a \#1 glass coverslip and spun at 3 krpm for 1 minute. The coverslips were exposed through a mask at 2100 \(\upmu\)W/cm\(^2\) for 40 seconds. Finally, they were developed in CD-26 for 15 seconds while stirring. Next, a titanium adhesion layer (20 nm) and a gold metallic layer (100 nm, 50 nm respectively) were deposited. After deposition, acetone was used to strip the resist. 

\subsection*{Temperature Setup}
Temperature measurements of the NVNDs' thermal properties were conducted using a heating stage supplied by Instec Inc (HCP621 V) rated for a temperature precision of 50 mK. 

\subsection*{GaN HEMT Preparation}
GaN HEMTs were aquired from Wolfspeed/Cree (CGHV1J006D-GP4) and were mounted on custom PCBs. Gate and drain voltages were controlled through programmable power supplies and a bias tee was used to feed in the 3 GHz frequency to drive the NV spins. A thermoelectric cooler mounted with a liquid cooling stage supplied by Koolance was used to cool the GaN HEMT's backside to 20 \(\degree\)C during operation. 

\section*{Materials and Correspondence}
The correspondence should be addressed to Dirk R. Englund (englund@mit.edu)  

\section{Data Availability}
The data that support the findings of this study are available from the corresponding authors on reasonable request.

\section{Acknowledgments}
This research is supported in part by the Army Research Office Multidisciplinary University Research Initiative (ARO MURI) biological transduction program. C.F. and M.W. acknowledges support from Master Dynamic Limited. M.T. acknowledges support by an appointment to the Intelligence Community Postdoctoral Research Fellowship Program at MIT, administered by Oak Ridge Institute for Science and Education through an interagency agreement between the U.S. Department of Energy and the Office of the Director of National Intelligence. E.N.W., L.Z., and K.B. acknowledge  funding  support  provided  by the  MIT/MTL  GaN  Energy  Initiative  and  the  Singapore-MIT Alliance for Research and Technology (SMART) LEES Program. K.B. also  acknowledges that  this  research  was  conducted  with  government  support under  and  awarded  by  DoD,  Air  Force  Office  of  Scientific  Research,  National  Defense  Science  and  Engineering Graduate (NDSEG) Fellowship, 32 CFR 168a. M.W. was supported in part by the STC Center for Integrated Quantum Materials (CIQM), NSF Grant No. DMR-1231319, in part by the Army Research Laboratory Center for Distributed Quantum Information (CDQI). L.Z. also acknowledges the funding support provided by the Office of Naval Research (ONR) under Award No. N00014-17-1-2363. The authors would also like to thank Sinan Karaveli, Tim Schr\"{o}der, Donggyu Kim, Alberto Lauri, Sara Mouradian, Eric Bersin, Kurt Broderick and David Lewis for their helpful discussion.

\bibliographystyle{unsrtnat}
\bibliography{References}

\end{document}